\documentstyle[prl,aps,epsf]{revtex}
\begin{document}
\draft
\twocolumn[\hsize\textwidth\columnwidth\hsize\csname @twocolumnfalse\endcsname

\title{Transverse Depinning in Strongly Driven Vortex Lattices with Disorder} 
\author{C.J.~Olson and C.~Reichhardt}
\address{Department of Physics, University of California, Davis, CA
95616}

\date{\today}
\maketitle
\begin{abstract}
Using numerical simulations we investigate the transverse 
depinning of moving vortex lattices interacting with random disorder. 
We observe a finite transverse depinning barrier for 
vortex lattices that are driven with high longitudinal  
drives, 
when the vortex lattice is defect free and moving 
in correlated 1D channels. 
The transverse barrier is reduced
as the longitudinal drive is decreased and defects appear
in the vortex lattice, and 
the barrier disappears 
in the plastic flow regime. 
At the transverse depinning transition, the vortex lattice 
moves in a staircase pattern with a clear transverse  
narrow-band voltage noise signature.  
\end{abstract}
\pacs{PACS numbers: 74.60.Ge}

%\vskip2pc
\vspace{-0.1in}
\vskip2pc]

The dynamics of driven vortex lattices interacting with disorder 
exhibit a wide variety of interesting nonequilibrium behavior and
dynamic phase transitions. 
Experiments 
\cite{Higgins1,Yaron2,Beasly3,Andrei4,Duarte5,Kes6,Pardo7,Troya8}, 
simulations \cite{Koshelev9,Moon10,Ryu11,Melt12,Olson13,Kolton14a}, 
and theory \cite{Koshelev9,Giamarchi14,Brag15,Balents16,Scheidl17} 
suggest that at low drives the vortex
lattice is disordered and exhibits plastic or random flow while at 
higher drives the lattice can undergo a  
reordering transition and flow elastically.
In this highly driven state it was 
suggested by Koshelev and Vinokur \cite{Koshelev9} 
that the flux lattice forms a moving crystal. 
In subsequent theoretical 
work, Giamarchi and Le Doussal \cite{Giamarchi14,Brag15} 
proposed that the reordered
state is actually an ordered 
moving glass phase and that the vortices travel
in highly correlated static channels. 
In other work \cite{Balents16,Scheidl17}, it has been proposed that these 
channels may be decoupled, producing a smectic 
structure.
Simulations \cite{Moon10,Ryu11,Olson13,Kolton14a} 
and experiments \cite{Pardo7} have found 
evidence for both smectic as well as more ordered moving vortex 
lattice structures. 

A particularly intriguing prediction of
the theory of Giamarchi and Le Doussal is that, in the 
highly driven phase,  
the moving lattice has a diverging potential barrier against 
a transverse driving force, resulting in  the existence of a 
{\it finite transverse critical current}.   
This transverse critical current has been observed in simulations
by Moon {\it et al.} \cite{Moon10} 
and Ryu {\it et al.}  \cite{Ryu11} in the highly driven 
phase.  
Large transverse barriers have also been seen in systems
containing periodic pinning \cite{Periodic18}.
Also, recent experiments \cite{Troya8} involving STM  
images of moving vortices reveal that the vortex lattice 
moves along one of its principle axes rather than 
in the direction of the drive (as predicted by \cite{Schmid19}), 
suggesting that the
moving lattice is stable to a 
small transverse force component. 
 
Although the existence of a  
transverse critical current has  been confirmed in simulations,
there has been no numerical study of the properties of the critical
current, such as the dependence of the barrier size on the strength
of the longitudinal drive or on the defectiveness of the vortex lattice.
It would also be very interesting to understand the {\it dynamics} of  
the vortices at the transverse depinning
transition, and relate this to experimental measures such as 
voltage noise spectra.

In this work we report a simulation
study of the transverse depinning transition
in driven vortex lattices interacting with random disorder.
We find that at high longitudinal 
drives, when the vortex lattice is 
defect free and moves in correlated 1D channels along one of its 
principle axes, a finite transverse 
depinning barrier is present. 
For lower drives the transverse barrier is reduced but still present, 
even in the decoupled channel limit when 
adjacent channels slip past each other and
some defects in 
the vortex lattice appear.
For the lowest drives the flux lattice 
becomes highly defected as it enters the plastic flow
phase, and the transverse barrier is lost.  
In the high driving limit at the transverse depinning transition,
the vortex lattice spends most of its time moving
along the longitudinal direction,
but periodically jumps in the transverse
direction by 
one lattice constant. The vortex lattice can thus be seen to move 
in a staircase like fashion, 
keeping its principle axis aligned in the original direction of 
longitudinal driving. 
As the transverse force is increased the frequency of the jumps
in the transverse direction increases. This motion 
produces a clear washboard signal in the transverse velocity which 
can be detected for transverse 
drives up to ten times the transverse depinning threshold. 

We consider a 2D slice of a system of superconducting
vortices interacting with a random 
pinning background.
The applied magnetic field ${\bf H}=H{\bf{\hat z}}$ is perpendicular
to our sample, and we use
periodic boundary conditions in $x$ and $y$. 
The $T=0$ overdamped equation of motion for a vortex is:
\begin{equation}
{\bf f}_{i} = \eta{\bf v}_{i} = {\bf f}_{i}^{vv} + {\bf f}_{i}^{vp} + 
{\bf f}_{d} +  {\bf f}_{i}^{T} \ ,
\end{equation}
where ${\bf f}_{i}$ is
the total force acting on vortex $i$, ${\bf v}_{i}$ is
the velocity of vortex $i$, and $\eta$ is the
damping coefficient, which is set to 1. 
The repulsive vortex-vortex interaction 
is given by
${\bf f}_{i}^{vv} = 
\sum_{j=1}^{N_{v}}A_{v}f_{0}K_{1}(|{\bf r}_{i} -{\bf r}_{j}|/
\lambda){\bf {\hat r}}_{ij}$ where 
${\bf r}_{i}$ is the position of vortex $i$,
$\lambda$ is the penetration depth,
$f_0=\Phi_{0}^{2}/8\pi\lambda^{3}$, 
the prefactor $A_v$ is set to 3 \cite{Olson13}, and 
$K_{1}(r/\lambda)$ is a modified Bessel function which falls off exponentially
for $r > \lambda$, allowing a cutoff in the interactions to be placed 
at $r = 6\lambda$ for computational efficiency. 
We use a vortex density of $n_v = 0.75/\lambda^2$ giving the number of vortices
$N_v=864$ for a sample of size $36\lambda \times 36\lambda$.
The pinning is modeled as 
randomly placed attractive parabolic traps of radius $r_p=0.3\lambda$ with
${\bf f}_{i}^{vp} = (f_{p}/r_{p})(|{\bf r}_{i} - {\bf r}_{k}^{(p)}
|)\Theta(r_{p} - |{\bf r}_{i} - {\bf r}_{k}^{(p)}|)
{\bf {\hat r}}^{(p)}_{ik}$,
where 
${\bf r}^{(p)}_{k}$ is the location of pin $k$,
$\Theta$ is the Heaviside step function, 
${\bf {\hat r}}_{ij} = ({\bf r}_{i} - {\bf r}_{j})/|{\bf r}_{i} - {\bf r}_{j}|$ 
and ${\bf {\hat r}}^{(p)}_{ik} = ({\bf r}_{i} - 
{\bf r}^{(p)}_{k})/|{\bf r}_{i} - {\bf r}^{(p)}_{k}|$.
The pin density is $n_p = 1.0/\lambda^2$ and the pinning force is 
$f_p = 1.5f_0$.
The Lorentz force from an applied current ${\bf J}=J{\bf {\hat y}}$
is modeled as a uniform 
driving force ${\bf f}_{d}$ on the vortices
in the $x$-direction. 
We initialize the vortex positions by performing 
simulated
annealing with $f_{d}/f_{0} = 0.0$. We then gradually increase $f_{d}$ to 
its final value by repeatedly   
increasing $f_{d}$ by $0.004f_{0}$ and remaining 
at each drive for $10^{4}$ time steps, where $dt = 0.02$.
If we increase the drive more rapidly than this, 
the reordered 
vortex lattice that forms at higher drives may fail to 
align its principle axis in the
direction of the driving. Slow increases in $f_{d}$ always produce an 
aligned lattice. 
Once the final $f_{d}$ value is reached we 
equilibrate  the system for
an additional $2\times 10^{4}$ steps and then begin applying a 
force in the transverse direction $f_{d}^{y}$ which we increase by 
$ 0.0001f_0$ every $10^{4}$ time steps. We monitor the transverse 
velocities 
$V_{y} = (1/N_{v})\sum^{N_{v}}_{i=1}{\bf v}_{i}\cdot {\bf {\hat y}}$
to identify the transverse critical current. 
  
In Fig.~1 we show $V_{y}$ versus the transverse drive $f_{d}^{y}$ 
at longitudinal drives of
$f_{d}/f_{0} = 1.0$ and $f_{d}/f_{0}=3.0$
for a system  
with a longitudinal depinning threshold of $f_{c}^{x}/f_0 \approx 0.5$.  
For $f_{d}/f_{0} = 3.0$ the vortex lattice
is free of defects and the vortices move in well defined 1D channels as 
seen from the vortex trajectories in the right inset, 
in agreement with previous simulations \cite{Olson13}. 
In this case there is clear evidence for a transverse barrier
with $f_{c}^{y}/f_{c}^{x} \approx 0.01$, approximately 100 times smaller than 
the longitudinal depinning threshold, in agreement with earlier 
simulations \cite{Moon10,Ryu11}. 
Thus, the vortex lattice resists changing its direction of motion.
For $f_{d}/f_{0} = 1.0$, the vortex lattice is highly defected and 
the 1D channel structure is lost (as seen in the 
left inset of Fig.~1).
In this case
the transverse barrier is 
absent since the lattice has no particular alignment and
can readily  change the direction of its motion. 
In the absence of pinning, $f_{c}^{y}=0$ for all drives $f_d$,
as indicated by the top curve in Fig.~1.

We ran a series of simulations in which the 
final longitudinal drive $f_d$ was varied
in order to determine 
the dependence of the magnitude of the transverse barrier on the
magnitude of the longitudinal
drive, as well as on the density of defects in the vortex lattice.
In Fig.~2 we plot the resulting transverse depinning thresholds 
$f_{c}^{y}$  and the 
fraction of six-fold coordinated vortices $P_{6}$ 
(calculated from the Voronoi or Wigner-Seitz cell construction) as 
a function of $f_{d}$. 
For longitudinal drives $f_{d}/f_0 > 1.5$, there are no defects in the 
vortex lattice (indicated by the fact that $P_{6} \approx 1.0$) 
and $f_{c}^{y}$ is roughly constant, $f_{c}^{y}/f_{c}^{x}\approx 0.01$. 
Below $f_{d}/f_0 \lesssim 1.5$  
defects begin to appear in the vortex lattice as 
adjacent moving channels decouple, and the overall vortex lattice 
develops a moving smectic structure. \cite{Olson13}. 
The transverse critical current $f_{c}^{y}$,  
which is still finite in this phase,  
becomes progressively reduced as more defects are generated. 
At the lowest drives,
$f_{d}/f_0 \gtrsim 1.0$, the transverse critical force is lost when  
the 1D channels are completely destroyed 
and the vortex lattice enters the amorphous 
plastic flow phase shown in the left inset of Fig.~1.
The dislocations in the lattice, which were aligned perpendicular to the
vortex motion at $f_d/f_0>1.0$, become randomly aligned at the transition
to plastic flow.
The loss of $f_{c}^{y}$ thus coincides with the loss of alignment
of the defects 
and the destruction of the 1D channels.

We have also checked the effect of finite system size 
on the magnitude of the transverse critical force
for $f_{d}/f_0 = 1.5, 2.0$ and $3.0$ for different system sizes 
($L = 24\lambda$, $36\lambda$, $48\lambda$ and $60\lambda$). 
In the inset of Fig.~2, we show that $f_{c}^{y}/f_{c}^{x}$ 
is not affected by the system size.
These results support the idea that it is the presence of defects 
in the lattice that reduce or destroy the transverse barrier,
rather than any type of matching effect with the system size,
and that 
as long as some form of channeling occurs the barrier will still be
present.

In order to view the dynamics of  
the vortex lattice at the transverse depinning
threshold we plot in Fig.~3 the vortex positions and trajectories
for the same system 
shown in Fig.~1 with 
$f_{d}/f_{0} = 3.0$ and $f_{d}^{y}/f_{c}^{x} = 0.011$. For clarity we 
have highlighted a particular row of vortices. 
In Fig.~3(a) the principle axis of the vortex lattice  
is aligned with the direction of the drive and the ordered lattice
is moving along this axis.
In Fig.~3(b)  
the entire vortex lattice has  translated by one lattice constant 
in the transverse direction. During the transition the lattice
moves at an angle to the longitudinal drive, following a different
axis of the lattice. 
Once the vortices have moved one lattice constant transverse to the drive, they
begin 
moving along the 
{\it same} channels that formed before the transverse translation. 
After this, the vortices move 
along the longitudinal direction
once again, as in Fig.~3(c), before jumping by another 
lattice constant in the transverse direction.  
At the transverse depinning transition, the vortex lattice 
thus moves in a staircase like manner, always keeping its 
principle axis aligned in the 
direction of the original longitudinal drive and always 
translating along one of the axes of the lattice. 
As $f_{c}^{y}$ is increased the frequency of jumps in the transverse 
direction increases. If  $f_{c}^{y}$ 
is increased to a high enough value,
we have found evidence that
the vortex lattice will reorient itself with the net driving force 
via the creation of a grain boundary.
This will be discussed in more detail elsewhere.
The transverse depinning transition is  
unlike the longitudinal depinning transition 
in that the latter occurs through plastic
deformations of the lattice 
and the generation of a large number of 
defects. In contrast, the transverse depinning 
transition is {\it elastic}.   

The fact that the vortices move periodically by a lattice constant in
the transverse direction is a result of the fact that the 
longitudinal channels followed by the vortices
are uniquely determined by the underlying disorder \cite{Giamarchi14}.  
The vortices
jump from one of these stable channels to another, giving the same
effect as a washboard potential.  This periodic effect occurs only
for a moving lattice in which 1D channels have formed; to a 
stationary lattice, the disorder would appear random.

A consequence of the staircase-like vortex motion 
just above the transverse 
depinning threshold 
is that the net transverse vortex velocity at a fixed $f_{d}^{y}$ should
show a clear washboard frequency which should increase for 
increasing $f_{d}^{y}$. 
In Fig.~4(a) we plot $V_{y}$ for samples with $f_{d}^{y}$ held fixed
at several different values just above the transverse depinning threshold
for a system with $f_d/f_0=3.0$. 
For $f_{d}^{y}/f_{c}^{y} = 0.011$ the $V_{y}$ shows periodic pulses
which correspond to the correlated transverse jumps of the vortex lattice
seen in Fig.~3. 
The flat portions of the voltage signal correspond to time periods when
the lattice is moving only in the longitudinal direction, between hops.
For increasing transverse  
drive the frequency of these pulses
also increases. 
The additional structure in the $V_y$ voltage pulses at lower values
of $f_{d}^{y}$ is characteristic of the underlying pinning, and varies
for different disorder realizations.  It occurs when the vortex lattice
moves slightly unevenly, with a small wobble, but no defects or tearing
occur in the lattice. The main feature of large
periodic pulses is always observed, and the wobble dies away at larger
transverse drives.
In Fig.~4(b) we show 
that the Fourier transform of 
the velocity signal $V_y$ for a driving force of
$f_{d}^{y}/f_{c}^{y} = 0.016$ exhibits a resonance frequency
at $\nu=6.0\times 10^{-5}$ inverse MD steps. In Fig.~4(c), 
the resonant frequency increases linearly with $f_{d}^{y}$.
We find that this resonance
persists for $f_{d}^{y}$ up to ten times larger then the 
transverse depinning threshold.
It should be possible to detect 
this washboard frequency
with Hall-noise measurements.

In recent experiments employing an STM to directly image a slowly moving
vortex lattice \cite{KesPrivate}, evidence for staircase-like
motion of the flux lattice has been observed.  In these experiments
the direction of the driving force could not be directly controlled,
but was assumed to be at a slight angle with respect to the principle
vortex lattice vector, so that a transverse component of the driving
force was present.  Further experiments in which the magnitude and
direction of the drive can be directly controlled are needed; however,
experimental imaging techniques such as STM or Lorentz microscopy
\cite{Tonomura} seem highly promising.

In summary we have investigated the transverse depinning of moving 
vortex lattices interacting with random disorder. We find that for 
high longitudinal drives where the vortex lattice is defect free a 
finite transverse barrier forms. For lower drives where defects in 
the vortex lattice form   
and the vortex lattice
has a smectic structure the transverse barrier is 
reduced but still finite. In the highly disordered plastic flow phase the
transverse barrier is absent.  The transverse depinning transition is
elastic, unlike the plastic longitudinal depinning transition, and
near this transition the vortex lattice moves in a staircase-like fashion.
We observe a washboard frequency in the transverse voltage signal
which can be detected for transverse drives up to ten times the
depinning drive.

We thank T. Giamarchi, P. Kes, P. Le Doussal, F. Nori, R. Scalettar,
and G. Zim{\' a}nyi for helpful discussions.
We acknowledge support from CLC and CULAR, administered by the
University of California.

\begin{figure}
\center{
\epsfxsize=3.5in
\epsfbox{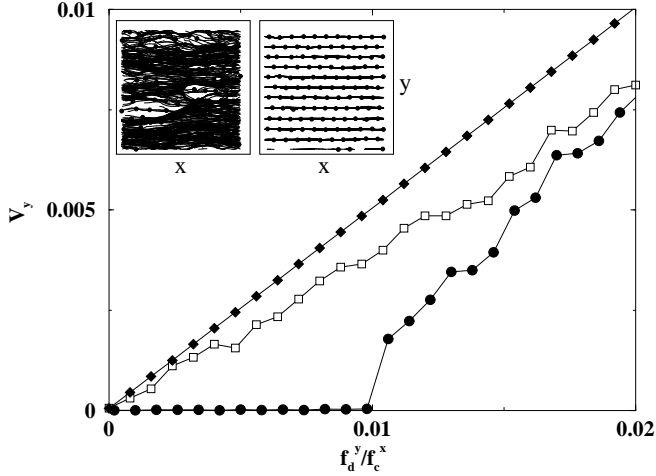}}
\caption{Transverse voltage $V_y$ versus transverse driving force
$f_{d}^{y}$.  Diamonds: A sample with no pinning, $f_{p}/f_{0}=0$,
shows no transverse barrier at any drive (here, $f_d/f_0=3.0$).  
Squares and circles:
a sample with pinning strength $f_{p}/f_{0}=1.5$ 
and longitudinal critical current $f_{c}^{x}/f_{0}=0.5$.  Squares:
For a longitudinal driving force $f_d/f_0= 1.0$ the vortices flow
plastically (left inset) and there is no transverse barrier.
Circles: 
For a longitudinal driving force $f_d/f_0= 3.0$ the vortices flow
in well-defined channels (right inset) and a transverse barrier
of $f_{c}^{y}/f_{c}^{x} = 0.01$ appears.  Insets:  Filled circles represent
vortices and lines show the paths traveled by the vortices while
moving over randomly spaced pinning sites (not shown).  Left inset: plastic
flow; right inset: channel flow.}
\end{figure}

\begin{figure}
\center{
\epsfxsize=3.5in
\epsfbox{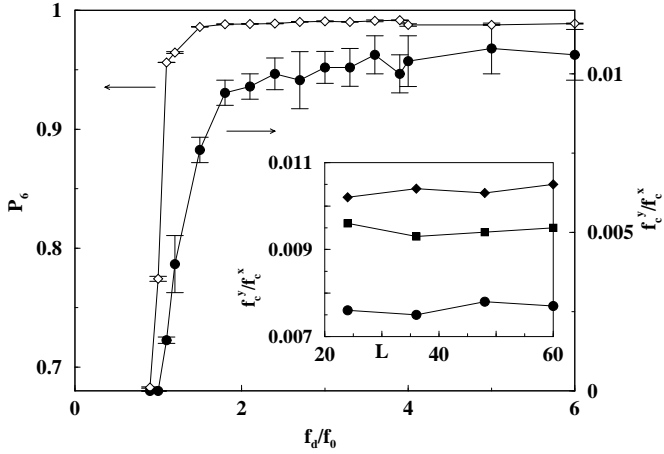}}
\caption{Circles: Transverse critical force $f_{c}^{y}/f_{c}^{x}$ versus
longitudinal driving force $f_d$.  Diamonds: Corresponding fraction of
six-fold coordinated vortices, $P_6$.  The transverse critical force
saturates for a reordered vortex lattice, $f_d/f_0 > 1.8$,
but drops as the
lattice becomes defected,  reaching zero at $f_d/f_0\approx 1.0$.
Inset: Transverse critical force for different
system sizes $L=24\lambda$, $36\lambda$, $48\lambda$, and $60\lambda$.
Circles: $f_d/f_0 = 1.5$; Squares: $f_d/f_0 = 2.0$; Diamonds: 
$f_d/f_0 = 3.0$.}
\end{figure}

\begin{figure}
\center{
\epsfxsize=3.5in
\epsfbox{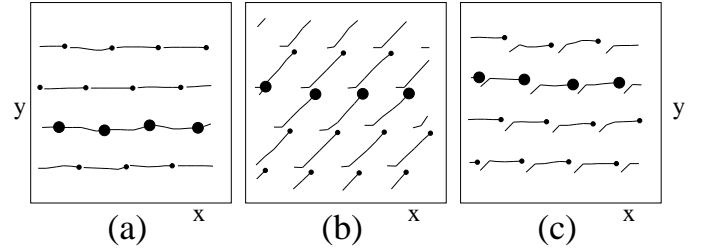}}
\caption{Consecutive
simulation images of vortex motion just above the transverse
critical force, $f_{d}^{y}/f_{c}^{y} = 0.011$.  Circles represent vortices
and lines indicate paths followed by the vortices. A particular row
of vortices has been highlighted.  In (a) the vortices are moving in
the direction of the applied longitudinal drive.  In (b) the vortex
lattice changes its direction of motion and follows a different
lattice vector until it has translated by one lattice constant.  In (c)
the lattice again switches direction and continues to flow in the 
longitudinal channels.  The frequency at which the vortex lattice hops
from channel to channel increases with increasing transverse drive.
}
\end{figure}

\begin{figure}
\center{
\epsfxsize=3.5in
\epsfbox{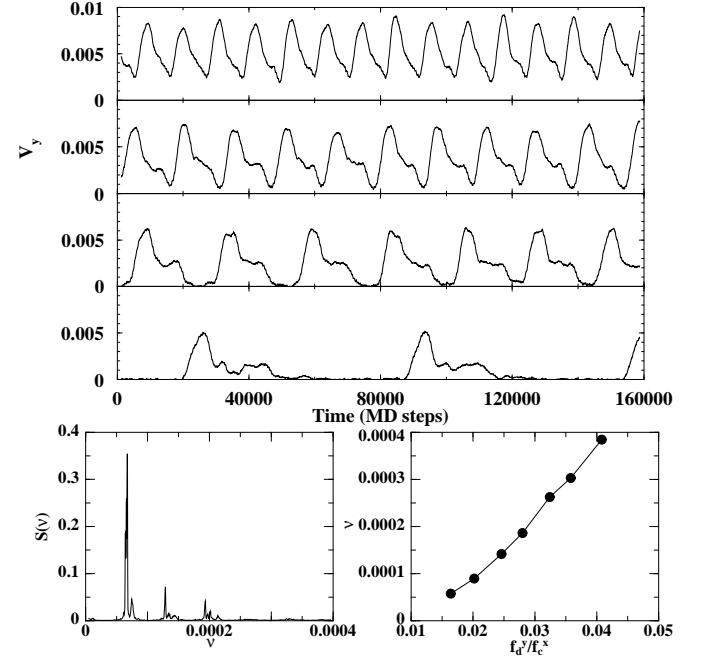}}
\caption{(a) Transverse voltage signal $V_y$ as a function of time
for a sample with longitudinal drive $f_d/f_0=3.0$ at
four different transverse driving currents $f_{d}^{y}$.  Bottom
to top: $f_{d}^{y}/f_{c}^{x}=0.011$, 
$f_{d}^{y}/f_{c}^{x}=0.014$, $f_{d}^{y}/f_{c}^{x}=0.016$, 
and $f_{d}^{y}/f_{c}^{x}=0.20$.
Each pulse corresponds to the vortex lattice moving one lattice
constant in the transverse direction.
(b) Power spectrum $S(\nu)$ of the transverse voltage noise signal 
$V_y$ for a sample
with $f_d/f_0=3.0$ and $f_{d}^{y}/f_{c}^{x}=0.016$.  
A clear narrow band signature appears at $\nu=6.0\times 10^{-5}$.
(c) Location of the narrow band peak $\nu$ for different
transverse driving forces $f_{d}^{y}$.
}
\end{figure}

\end{document}